\documentclass[12pt]{article}
\begin{document}
\setcounter{page}{1}
\def\theequation{\arabic{section}.\arabic{equation}}
\def\theequation{\thesection.\arabic{equation}}
\setcounter{section}{0}

\title{On the Solar Neutrino Problems, SNO experimental data and
low--energy nuclear forces}

\author{A. N. Ivanov~\thanks{E--mail: ivanov@kph.tuwien.ac.at, Tel.:
+43--1--58801--14261, Fax: +43--1--58801--14299}~${
^\ddagger}$, H. Oberhummer~\thanks{E--mail: ohu@kph.tuwien.ac.at,
Tel.: +43--1--58801--14251, Fax: +43--1--58801--14299} ,
N. I. Troitskaya~\thanks{Permanent Address: State Technical
University, Department of Nuclear Physics, 195251 St. Petersburg,
Russian Federation}}

\date{\today}

\maketitle

\begin{center}
{\it Institut f\"ur Kernphysik, Technische Universit\"at Wien,\\
Wiedner Hauptstr. 8-10, A-1040 Vienna, Austria}
\end{center}

\begin{center}
\begin{abstract}
The Solar Neutrino Problems (SNP's) are analysed within the Standard
Solar Model (BP2000) supplemented by the reduction of the solar
neutrino fluxes through the decrease of the solar core temperature.
The former can be realized through the enhancement of the
astrophysical factor for solar proton burning. The enhancement, the
upper bound of which is restricted by the helioseismological data,
goes dynamically due to low--energy nuclear forces described at the
quantum field theoretic level. The agreement of the reduced solar
neutrino fluxes with the experimental data is obtained within the
scenario of vacuum two--flavour neutrino oscillations. We show that by
fitting the mean value of the solar neutrino flux measured by
HOMESTAKE Collaboration we predict the high energy solar neutrino flux
measured by SNO Collaboration $\Phi^{\rm SNO}_{\rm th}({^8}{\rm B}) =
1.84\times 10^6\,{\rm cm^{-2}\,s^{-1}}$ in perfect agreement with
experimental value $\Phi^{\rm SNO}_{\exp}({^8}{\rm B}) = (1.75\pm
0.14)\times 10^6\,{\rm cm^{-2}\,s^{-1}}$ obtained via the measurement
of the rate of reaction $\nu_{\rm e}$ + D $\to$ + p + p + e$^-$
produced by ${^8}{\rm B}$ solar neutrinos. The theoretical flux for
low--energy neutrino flux measured by GALLIUM (GALLEX, GNO and SAGE)
Collaborations $S^{\rm Ga}_{\rm th}= 65\,{\rm SNU}$ agrees with the
experimental data averaged over experiments $S^{\rm Ga}_{\exp} =
(75.6\pm 4.8)\,{\rm SNU}$.
\end{abstract}
\end{center}

\begin{center}
\end{center}

\newpage

\section{Solar neutrino fluxes. Theory and Experiment}
\setcounter{equation}{0}

\hspace{0.2in} The Solar Neutrino Problem (SNP) [1,2] as the
disagreement with the theoretical prediction and the seminal
experimental data by Raymond Davis [3] has been recently reformulated
by Bahcall in the form of Three Solar Neutrino Problems [4]. The
solution of these SNP's demands simultaneous description of the
experimental data by HOMESTAKE, GALLEX--GNO--SAGE,
SUPERKAMIOKANDE and SNO Collaborations [4].

Nowadays there is no doubts that the solution of the SNP's goes via
the application of a mechanism of neutrino oscillations introduced in
physics by Gribov and Pontecorvo [5,6]. According to
Gribov--Pontecorvo's hypothesis electronic neutrinos $\nu_{\rm e}$
produced in the solar core can change their flavour due to the
transition $\nu_{\rm e} \to \nu_{\mu}$ during their travel to the
Earth. The former should obviously diminish the flux of solar
electronic neutrinos measured on the Earth.

The solar core can interfere in the process of neutrino oscillations in
a twofold way. First, neutrino oscillations can be resonantly
enhanced by virtue of the solar core matter suggested by Wolfenstein,
Mikheyev and Smirnov [7], so--called the MSW effect [8], and,
secondary, due to low--energy nuclear forces contributing to nuclear
reaction on $\nu_{\rm e}$--neutrino production. The most popular is
the MSW effect, since it allows to diminish the solar neutrino fluxes
without change of the main parameters of the Standard Solar Model
(SSM BP2000) formulated by Bahcall with co--workers [9,10]. 

The main nuclear reaction in the Sun is the solar proton burning p + p
$\to$ D + e$^+$ + $\nu_{\rm e}$. It is induced by the charge weak
current, that is defined by the $W^+$--boson exchange, and strong
low--energy nuclear forces.  The reaction p + p $\to$ D + e$^+$ +
$\nu_{\rm e}$ gives start the p--p chain of nucleosynthesis in the
Sun and main--sequence stars [2,11]. In the SSM the total (or
bolometric) luminosity of the Sun $L_{\odot} = (3.846\pm 0.008)\times
10^{26}\,{\rm W}$ is normalized to the astrophysical factor $S_{\rm
pp}(0)$ for the solar proton burning. The recommended value $S^{\rm
SSM}_{\rm pp}(0) = 4.00\times 10^{-25}\,{\rm MeV b}$ [12]. The
helioseismological data restrict the value of the astrophysical factor
$S_{\rm pp}(0)$ and predict $0.94 \le S_{\rm pp}(0)/S^{\rm SSM}_{\rm
pp}(0) \le 1.18$ [13].

The interference of low--energy nuclear forces into neutrino
production can be realized in the form of an enhancement of the
astrophysical factor $S_{\rm pp}(0)$. As has been recently shown
within both the Nambu--Jona--Lasinio model of light nuclei (the NNJL
model) [14-16] and the Effective Field Theory (EFT) [17--19] the
astrophysical factor $S_{\rm pp}(0)$ calculated within quantum field
theoretic models contains an arbitrary parameter which can be fixed
from the experimental data on the reactions for disintegration of the
deuteron by neutrinos and antineutrinos [16--19]. For example, in the
NNJL model the astrophysical factor $S_{\rm pp}(0)$ is defined by
[15,16]
\begin{eqnarray}\label{label1.1}
S_{\rm pp}(0) = (1 + \bar{\xi}\,)^2\times 4.08\times 10^{-25}\,{\rm
MeV\,b},
\end{eqnarray}
where $\bar{\xi}$ is an arbitrary parameter (see Appendix of
Ref.\,[15]) and Ref.\,[16]. The same factor appears in the cross
sections for neutrino and antineutrino disintegration of the deuteron
[15,16]. For example, for the cross sections for the disintegration of
the deuteron by reactor antineutrinos averaged over the antineutrino
energy spectrum we have [15]
\begin{eqnarray}\label{label1.2}
\langle\sigma^{\rm \bar{\nu}_{\rm e} D \to e^+nn}(E_{\bar{\nu}_{\rm
 e}})\rangle &=&(1+\bar{\xi})^2\times 11.56\times 10^{-45}\,{\rm
 cm}^2,\nonumber\\ \langle\sigma^{\rm \bar{\nu}_{\rm e}D \to
 \bar{\nu}_{\rm e}np}(E_{\bar{\nu}_{\rm e}})\rangle &=&
 (1+\bar{\xi})^2\times 6.28\times 10^{-45}\,{\rm cm}^2,
\end{eqnarray}
The experimental values of these cross sections read [20]
\begin{eqnarray}\label{label1.3}
 \langle\sigma^{\rm \bar{\nu}_{\rm e} D \to e^+nn}(E_{\bar{\nu}_{\rm
e}})\rangle_{\exp}&=& (9.83\pm 2.04)\times 10^{-45}\,{\rm
cm}^2,\nonumber\\ \langle\sigma^{\rm \bar{\nu}_{\rm e}D \to
\bar{\nu}_{\rm e}np}(E_{\bar{\nu}_{\rm e}})\rangle_{\exp} &=& (6.08\pm
0.77)\times 10^{-45}\,{\rm cm}^2.
\end{eqnarray}
We would like to accentuate that the averaged value of the cross
section for the reaction $\bar{\nu}_{\rm e}$ + D $\to$ n + n + e$^+$
has been calculated without account for reactor antineutrino
oscillations [21--23] which should diminish the theoretical values of
the cross section [22].

In Ref.\,[16] there has been a suggestion to use the ambiguity in the
calculation of the astrophysical factor in order to enhance its value.
This should lead to the decrease of the solar core
temperature. Indeed, any change of the astrophysical factor $S_{\rm
pp}(0)$ entails the change of the solar core temperature [24]:
\begin{eqnarray}\label{label1.4}
\frac{\Delta T_c}{T^{\rm SSM}_c} = -\,0.15\,\frac{\Delta S_{\rm
pp}(0)}{S^{\rm SSM}_{\rm pp}(0)}.
\end{eqnarray}
The enhancement of the astrophysical factor relative to the standard
value $S^{\rm SSM}_{\rm pp}(0) = 4.00\times 10^{-25}\,{\rm MeV\,b}$,
i.e. $\Delta S_{\rm pp}(0) > 0$, provides the decrease of the solar
core temperature. The maximal decrease of $T_c$ is restricted from
above by the inequality $\Delta S_{\rm pp}(0) \le 0.18\,S^{\rm
SSM}_{\rm pp}(0)$ that has been pointed out by Degl'Innocenti,
Fiorentini and Ricci [13]. Hence, the minimal value of the solar core
temperature, calculated for $T^{\rm SSM}_c = 1.57\times 10^7\,{\rm K}$
[10], can be equal to
\begin{eqnarray}\label{label1.5}
T_c = 1.53\times 10^7\,{\rm K}.
\end{eqnarray}
It is well--known that solar neutrino fluxes are sensitive to the value
of the solar core temperature [25]. By using a temperature dependence
of the solar neutrino fluxes obtained by Bahcall and Ulmer [25]:
$\Phi({\rm pp}) \propto T^{-1.1}_c$, $\Phi({\rm pep}) \propto
T^{-2.4}_c$, $\Phi({^7}{\rm Be}) \propto T^{\,10}_c$, $\Phi({^8}{\rm
B}) \propto T^{\,24}_c$, $\Phi({^{13}}{\rm N}) \propto T^{\,24.4}_c$
and $\Phi({^{15}}{\rm O}) \propto T^{\,27.1}_c$ we can calculate the
solar neutrino fluxes for the reduced solar core temperature
Eq.(\ref{label1.5}). The new values of the solar neutrino fluxes are
given in Table 3.

It is seen that the solar neutrino fluxes calculated for the solar
core temperature $T_c = 1.53\times 10^7\,$K are still not enough
decreased in order to satisfy the experimental data. Therefore, for
the reduction of the solar neutrino fluxes, taking place outside the
solar core, we will use the mechanism of neutrino oscillations. We
will follow the simplest scenario of vacuum two--flavour neutrino
oscillations suggested by Gribov and Pontecorvo [5,6]. By virtue of
the vacuum two--flavour neutrino oscillations $\nu_{\rm e} \to
\nu_{\mu}$ the solar neutrino fluxes should be multiplied by the
factor [5]
\begin{eqnarray}\label{label1.6}
P_{\nu_{\rm e} \to \nu_{\rm e}}(E_{\nu_{\rm e}}) = 1 -
\frac{1}{2}\,\sin^2 2\,\theta\,\Bigg(1 - \cos\frac{\Delta m^2
L}{2E_{\nu_{\rm e}}}\Bigg),
\end{eqnarray}
where $\Delta m^2 = m^2_{\nu_{\mu}} - m^2_{\nu_{\rm e}}$, $L$ is the
distance of the neutrino's travel, $E_{\nu_{\rm e}}$ is a neutrino
energy and $\theta$ is a neutrino--flavour mixing angle [5]. After
the averaging over energies and for $L$ of order of the Sun--Earth
distance the solar neutrino fluxes become multiplied by a factor [6]
\begin{eqnarray}\label{label1.7}
\overline{P_{\nu_{\rm e} \to \nu_{\rm e}}(E_{\nu_{\rm e}})} = 1 -
\frac{1}{2}\,\sin^2 2\,\theta.
\end{eqnarray}
The result of the integration over energies Eq.(\ref{label1.6}) can
occur only if $\Delta m^2 L/2 E_{\nu_{\rm e}}$ obeys the constraint
\begin{eqnarray}\label{label1.8}
\frac{\Delta m^2 L}{2 E_{\nu_{\rm e}}} \gg 1.
\end{eqnarray}
If we would like to get the factor Eq.(\ref{label1.7}) for all solar
neutrino fluxes including the ${^8}{\rm B}$ neutrinos, so that the
upper bound on the neutrino energies should coincide with the upper
bound on the ${^8}{\rm B}$ neutrino energy spectrum and should be
equal to $E_{\nu_{\rm e}} = 15\,{\rm MeV}$. As the Sun--Earth distance
$L$ amounts to $L = 1.496\times 10^{13}\,{\rm cm} = 7.581\times
10^{23}\,{\rm MeV^{-1}}$, the inequality Eq.(\ref{label1.7}) gives the
lower bound on $\Delta m^2$:
\begin{eqnarray}\label{label1.9}
\Delta m^2 \gg 4 \times 10^{-11}\,{\rm eV}^2.
\end{eqnarray}
The value of the mixing angle $\sin^2 2\theta$ we can get fitting, for
example, the mean value of the neutrino flux measured by HOMESTAKE
Collaboration. This gives 
\begin{eqnarray}\label{label1.10}
\sin^2 2\theta = 0.838.
\end{eqnarray}
The solar neutrino fluxes reduced by virtue of vacuum two--flavour
neutrino oscillations are adduced in Table 4. One can see a reasonable
agreement between theoretical and experimental values of the solar
neutrino fluxes for GALLEX--GNO and SAGE experiments. 

The theoretical expression for the solar neutrino flux $\Phi^{\rm
SNO}_{\rm th}({^8}{\rm B})$ measured by SNO Collaboration via the
measurement of the rate of reaction $\nu_{\rm e}$ + D $\to$ + p + p +
e$^-$ produced by ${^8}{\rm B}$ solar neutrinos can be defined by
\begin{eqnarray}\label{label1.11}
\Phi^{\rm SNO}_{\rm th}({^8}{\rm B}) = 1.18\,(1 -
0.5\,\sin^22\theta)\,\Phi({^8}{\rm B}) = (1.84^{+0.36}_{-0.26})\times
10^6\,{\rm cm^{-2}s^{-1}},
\end{eqnarray}
where the factors $(1 - 0.5\,\sin^22\theta)$ and 1.18 take into
account the contribution of vacuum two--flavour neutrino oscillations
and the enhancement of the cross section for the reaction $\nu_{\rm
e}$ + D $\to$ e$^-$ + p + p due to the enhancement of $S_{\rm pp}(0)$
by a factor $(1 + \bar{\xi})^2 = 1.18$ caused by invariance under time
reversal\,\footnote{Indeed, the experimentally measured solar neutrino
signal is proportional to the cross section for the reaction used for
the registration of solar neutrinos [2,26]. This means that the signal
increases, when the interaction of neutrinos with the target
enhances.}.  The theoretical value $\Phi^{\rm SNO}_{\rm th}({^8}{\rm
B}) = (1.84^{+0.36}_{-0.26}) \times 10^6\,{\rm cm^{-2}s^{-1}}$ is in
good agreement with recent experimental data by SNO Collaboration
$\Phi^{\rm SNO}_{\exp}({^8}{\rm B}) = (1.75\pm 0.14) \times 10^6\,{\rm
cm^{-2}s^{-1}}$ measured from the rate of the reaction $\nu_{\rm e}$ +
D $\to$ p + p + e$^-$ produced by ${^8}{\rm B}$ solar neutrinos (see
Table 1).

\section{Conclusion}
\setcounter{equation}{0}

\hspace{0.2in} We have suggested a reduction of the solar core
temperature due to a dynamics of low--energy nuclear forces described
at the quantum field theoretic level [14--19]. We have shown that the
reduction of the solar neutrino fluxes in the solar core caused by the
decrease of the solar core temperature supplemented by the scenario of
vacuum two--flavour neutrino oscillations $\nu_{\rm e} \leftrightarrow
\nu_{\mu}$ during the travel of solar neutrinos to the Earth proposed
by Gribov and Pontecorvo [5] gives a reasonable theoretical basis for
the understanding of the SNP's formulated by Bahcall [4].

It is important to emphasize that the necessary reduction of the solar
core temperature makes up only 2.7$\%$ of the temperature recommended
by SSM (BP2000) [10]
\begin{eqnarray*}
\Delta(T_c) = \frac{T^{\rm SSM}_c - T_c}{T^{\rm SSM}_c} = 2.7\,\%.
\end{eqnarray*}
This agrees with the constraints $\Delta(T_c) = \pm 6\%$ on the solar
core temperature fluctuations given by Bethe and Bahcall \cite{[BET]}.

Due to the reduction of the neutrino fluxes in the solar core for the
secondary reduction caused by two--flavour neutrino oscillations we
obtain $\Delta m^2 \gg 4\times 10^{-11}\,{\rm eV}^2$ and
$\sin^22\theta = 0.838$. This allows a simultaneous description of the
experimental data on the solar neutrino fluxes with an accuracy not
worse than the theoretical accuracy of the SSM (BP2000). The
constraint $\Delta m^2 \gg 4\times 10^{-11}\,{\rm eV}^2$ is rather
general and makes the analysis of two--flavour neutrino oscillations
much more flexible with respect to different experimental data on
parameters of neutrino--flavour oscillations \cite{[RAF]}.

The mixing angle $\sin^22\theta = 0.838$ has been obtained by fitting
the {\it mean value} of the experimental flux measured by
HOMESTAKE. The theoretical value of the low--energy solar neutrino
flux $S^{\rm Ga}_{\rm th}= 65\,{\rm SNU}$ agrees with experimental
data by GALLIUM Collaborations $S^{\rm Ga}_{\exp} = (75.6\pm
4.8)\,{\rm SNU}$ averaged over experiments.  Our theoretical
prediction for the ${^8}{\rm B}$ solar neutrino flux $\Phi^{\rm
SNO}_{\rm th}({^8}{\rm B}) = 1.84\times 10^6\,{\rm cm^{-2}\,s^{-1}}$,
that should be measured on the Earth, is in perfect agreement with the
experimental value $\Phi^{\rm SNO}_{\exp}({^8}{\rm B}) = (1.75\pm
0.14)\times 10^6\,{\rm cm^{-2}\,s^{-1}}$ obtained via the measurement
of the rate of reaction $\nu_{\rm e}$ + D $\to$ + p + p + e$^-$
produced by ${^8}{\rm B}$ solar neutrinos. The cross section for the
reaction $\nu_{\rm e}$ + D $\to$ + p + p + e$^-$ as well as the
astrophysical factor for the solar proton burning $S_{\rm pp}(0)$ is
induced by the weak charge current and defined by low--energy nuclear
forces, which are described well by the NNJL model in complete
agreement with nuclear phenomenology. For example, the D--wave
component of the wave function of the deuteron $D/S = 0.0238$,
calculated relative to the S--wave component without input parameters
(see EPJA{\bf 12}, 87 (2001) of Ref.[14]), is in agreement with the
phenomenological value $D/S = 0.0256\pm 0.0004$ used by Kamionkowski
and Bahcall for the description of the realistic wave function of the
deuteron in connection with the calculation of the astrophysical
factor $S_{\rm pp}(0)$ within the potential model approach
\cite{[BAH]}. The reaction $\nu_{\rm e}$ + D $\to$ + p + p + e$^-$ is
very sensitive to neutrino oscillations and reproduces the net rest of
the solar neutrino flux originated by the boron decay ${^8}{\rm B} \to
{^8}{\rm Be^*}$ + e$^+$ + $\nu_{\rm e}$ in the solar core. Since the
cross section for the reaction $\nu_{\rm e}$ + D $\to$ + p + p + e$^-$
is defined in the NNJL model by the same dynamics of strong
low--energy nuclear forces as the astrophysical factor $S_{\rm
pp}(0)$, the obtained agreement becomes not surprising. Such an
agreement testifies also the consistency of the dynamics of strong
low--energy nuclear forces described by the NNJL model as well as the
EFT with the SSM (BP2000) [10]. In fact, the neutrino fluxes decreased
by the change of the solar core temperature are fully determined by
the SSM (BP2000) and the temperature law--scaling suggested by the SSM
[25].

Thus, the suggested scenario of the evolution of solar neutrino fluxes
reconciles the experimental data on high energy solar neutrino fluxes
by HOMESTAKE, SNO and SUPERKAMIOKANDE Collaborations and low--energy
solar neutrino fluxes by GALLIUM Collaborations GALLEX, GNO and SAGE
with theoretical predictions and relaxes the stress of the SNP's
formulated by Bahcall [4].

Since the constraint $\Delta m^2 \gg 4\times 10^{-11}\,{\rm eV}^2$
means that effectively the theoretical values of the solar neutrino
fluxes do not depend on $\Delta m^2$, the agreement between
theoretical solar neutrino fluxes and experimental data is reached by
virtue of the tuning of only two parameters (i) the solar core
temperature $T_c$, diminished by 2.7$\%$ relative to the solar core
temperature recommended by SSM (BP2000) due to the dynamics of
low--energy nuclear forces and restricted from above by
helioseismological data, and (ii) the mixing angle $\sin^2\theta =
0.838$, fixed by the fit of the mean value of the solar neutrino flux
measured by HOMESTAKE Collaboration.

Now let us discuss the experimental data by SNO and SUPERKAMIOKANDE
obtained by the measurement of the ${^8}{\rm B}$ solar neutrino flux
using the reaction $\nu_{\rm x}$ + e$^-$ $\to $ e$^-$ + $\nu_{\rm x}$,
where $\nu_{\rm x} = \nu_{\rm e}, \nu_{\mu}$ or
$\nu_{\tau}$. Accounting for the contributions of electronic and
muonic neutrinos, $\nu_{\rm x} = \nu_{\rm e}$ and $\nu_{\mu}$, we
obtain the theoretical value of the neutrino flux equal to $\Phi^{\rm
\nu_{\rm e}e}_{\rm th}({^8}{\rm B}) = (1.73^{+0.34}_{-0.24})\times
10^6\,{\rm cm^{-2}\,s^{-1}}$ [16]. This value agrees within two
standard deviations with the data by SNO, $\Phi^{\rm \nu_{\rm
e}e}_{\exp}({^8}{\rm B})_{\rm SNO} = (2.39 \pm 0.34)\times 10^6\,{\rm
cm^{-2}\, s^{-1}}$, and qualitatively with the experimental data by
SUPERKAMIOKANDE, $\Phi^{\rm \nu_{\rm e}e}_{\exp}({^8}{\rm B})_{\rm SK}
= (2.32 \pm 0.09)\times 10^6\,{\rm cm^{-2}\, s^{-1}}$, (see Table
1). 

We have considered the simplest scenario of vacuum two--flavour
neutrino oscillations. The inclusion of the $\nu_{\tau}$ neutrino as
well as the {\it sterile} $\nu_{\rm s}$ neutrino in the scenario of
neutrino oscillations is implied by the experimental data on
atmospheric neutrinos by SUPERKAMIOKANDE [27]. However, the scenarios
of three--flavour neutrino oscillations $(\nu_{\rm e}, \nu_{\mu},
\nu_{\tau})$ or four--flavour neutrino oscillations, if to include
{\it sterile} neutrinos, are rather flexible due to a big number of
input parameters [28]. Therefore, one can hope that the results
calculated for the two--flavour neutrino oscillations can be left
practically unchanged.

The theoretical predictions for two--flavour neutrino--oscillation
parameters: $\Delta m^2 \gg 4\times 10^{-11}\,{\rm eV}^2$ and
$\sin^22\theta = 0.838$ should be applied to the calculation of the
contribution of reactor antineutrino oscillations [21--23] to the
cross sections for antineutrino disintegration of the deuteron
$\bar{\nu}_{\rm e}$ + D $\to$ n + n + e$^+$ and $\bar{\nu}_{\rm e}$ +
D $\to$ n + p + $\bar{\nu}_{\rm e}$ in order rearrange the theoretical
enhancement of the astrophysical factor $S_{\rm pp}(0)$ with
theoretical cross sections for the reactions $\bar{\nu}_{\rm e}$ + D
$\to$ n + n + e$^+$ and $\bar{\nu}_{\rm e}$ + D $\to$ n + p +
$\bar{\nu}_{\rm e}$ and experimental data of Reines's experimental
group [20]. We are planning to carry out this work in our forthcoming
publications.

\section*{Acknowledgement}

\hspace{0.2in} One of the authors (N. I. Troitskaya) is grateful to
the staff of Atomic and Nuclear Institute of the Austrian Universities
and especially to Manfried Faber for financial support and warm
hospitality extended to her during her stay at Vienna, when this work
was completed.

\newpage

\noindent Table 1.  Solar neutrino data, $1\,{\rm SNU} =
10^{-36}\,{\rm events/(atoms\cdot s)}$. The error is defined as
$\sigma = \sqrt{({\rm stat.})^2 +({\rm syst.})^2}$
\vspace{0.2in}

\begin{tabular}{|c|c|c|  }\hline
\cline{2-3}Experiment & Data $\pm \,\sigma$ & Units\\ \hline HOMESTAKE
\cite{[HOM]} & & \\ $\nu_{\rm e}$ + ${^{37}}{\rm Cl}$ $\to$ e$^-$ +
${^{37}}{\rm Ar}$&$ 2.56\pm 0.23$& SNU \\ $E_{\rm th} = 0.81\,{\rm
MeV}$ & & \\ \hline SAGE \cite{[SAGE]}& & \\ $\nu_{\rm e}$ +
${^{71}}{\rm Ga}$ $\to$ e$^-$ + ${^{71}}{\rm Ge}$ & $77.0\pm 6.7$ &
SNU \\ $E_{\rm th} = 0.23\,{\rm MeV}$ & & \\ \hline GALLEX + GNO
\cite{[GNO]}& & \\ $\nu_{\rm e}$ + ${^{71}}{\rm Ga}$ $\to$ e$^-$ +
${^{71}}{\rm Ge}$ & $74.1\pm 6.8$ & SNU\\ $E_{\rm th} = 0.23\,{\rm
MeV}$ & &\\ \hline KAMIOKANDE \cite{[KAM]}& & \\ $\nu_{\rm e}$ + e$^-$
$\to$ e$^-$ + $\nu_{\rm e}$ & $2.80\pm 0.38$ & $10^6\,{\rm
cm^{-\,2}s^{-\,1}}$\\ $E_{\rm th} = 7.0\,{\rm MeV}$ & &\\ \hline
SUPERKAMIOKANDE \cite{[SUP]}& & \\ $\nu$ + e$^-$ $\to$ $\nu$ + e$^-$ &
$2.32\pm 0.09$ & $10^6\,{\rm cm^{-2}s^{-1}}$\\ $E_{\rm th} = 5.5\,{\rm
MeV}$ & & \\ \hline SNO \cite{[SNO]}& & \\ $\nu$ + D $\to$ p + p +
e$^-$ & $1.75\pm 0.14$ & $10^6\,{\rm cm^{-\,2}s^{-\,1}}$\\ $E_{\rm th}
= 7.26\,{\rm MeV}$ & \\ \hline SNO \cite{[SNO]}& & \\ $\nu$ + e$^-$
$\to$ $\nu$ + e$^-$ & $2.39\pm 0.34$ & $10^6\,{\rm
cm^{-\,2}s^{-\,1}}$\\ $E_{\rm th} = 7.26\,{\rm MeV}$ & & \\ \hline
\end{tabular}\\
\vspace{0.2in}

\noindent Table 2. Standard Solar Model (BP2000) predictions for
the solar neutrino fluxes normalized to the recommended value of the
astrophysical factor $S_{\rm pp}(0) = 4.00\times 10^{-25}\,{\rm
MeV\,b}$ (see Table 7 of Ref.\,[10]).
\vspace{0.2in}

\begin{tabular}{|c|c|c|c|c|  }\hline
\cline{2-4} Source & Flux & Cl& Ga & SNO -- SK \\ & $(10^{10}\,{\rm
cm^{-2}s^{-1}})$ & (SNU)& (SNU)& $(10^6\,{\rm cm^{-2}s^{-1}})$\\
\hline pp & 5.95$(1.00\pm 0.01)$ & 0.0 & 69.7 & \\ pep & $1.40\times
10^{-2}(1.00\pm 0.015)$ & 0.22& 2.8 &\\ $^7{\rm Be}$ & $4.77\times
10^{-1}(1.00\pm 0.01)$ & 1.15 & 34.2 & \\ $^{8}{\rm B}$ & $5.05\times
10^{-4}(1.00^{+0.20}_{-0.16})$ & 5.76 & 12.1 &
$5.05(1.00^{+0.20}_{-0.16})$\\ $^{13}{\rm N}$ & $5.48\times
10^{-2}(1.00^{+0.21}_{-0.17})$ & 0.09 & 3.4 & \\ $^{15}{\rm O}$ &
$4.80\times 10^{-2}(1.00^{+0.25}_{-0.19})$ & 0.33 & 5.5& \\ \hline & &
$7.6^{+1.3}_{-1.1}$ & $128^{+9}_{-7}$ & $5.05^{+1.01}_{-0.81}$\\
\hline
\end{tabular}\\
\vspace{0.2in}

\newpage

\noindent Table 3. The solar neutrino
fluxes normalized to astrophysical factor $S_{\rm pp}(0) = 4.72\times
10^{-25}\,{\rm MeV\,b}$ caused by the non--trivial contribution of the
nucleon tensor current [15,16].

\vspace{0.2in}

\begin{tabular}{|c|c|c|c|c|  }\hline
\cline{2-4} Source & Flux & Cl& Ga & SNO -- SK \\ & $(10^{10}\,{\rm
cm^{-2}s^{-1}})$ & (SNU)& (SNU)& $(10^6\,{\rm cm^{-2}s^{-1}})$\\
\hline pp & 6.10 & 0.0 & 71.49 & \\ pep & $1.48\times 10^{-2}$ & 0.21&
2.98 &\\ $^7{\rm Be}$ & $3.66\times 10^{-1}$ & 0.88 & 26.23 & \\
$^{8}{\rm B}$ & $2.69\times 10^{-4}$ & 3.08 & 6.46 & 2.69\\ $^{13}{\rm
N}$ & $3.13\times 10^{-2}$ & 0.05 & 1.91& \\ $^{15}{\rm O}$ &
$2.56\times 10^{-2}$ & 0.19 & 2.89& \\ \hline & & $4.41$ & $111.96$ &
\\ \hline
\end{tabular}\\
\vspace{0.2in}

\noindent Table 4. The solar neutrino fluxes normalized to $S_{\rm
pp}(0)= 1.18\,S^{\rm SSM}_{\rm pp}(0) = 4.72\times 10^{-25}\,{\rm
MeV\,b}$. The theoretical values of experimentally measured neutrino
fluxes are calculated within a scenario of vacuum two--flavour
neutrino oscillations at $\sin^22\theta = 0.838$.  The error is
defined as $\sqrt{({\rm stat.})^2 +({\rm syst.})^2}$. The experimental
value of GALLIUM Collaborations is averaged over experimental data
(see Table 1 of Ref.\,[4], hep--ph/0108147).

\vspace{0.2in}

\begin{tabular}{|c|c|c|c|c|  }\hline
\cline{2-4} Source & Flux & Cl& Ga & SNO -- SK \\ & $(10^{10}\,{\rm
cm^{-2}s^{-1}})$ & (SNU)& (SNU)& $(10^6\,{\rm cm^{-2}s^{-1}})$\\
\hline pp & $6.10$ & 0.0 & $41.54$ & \\ pep & $1.48\times 10^{-2}$ &
$0.13$ &1.73 &\\ $^7{\rm Be}$ & $3.66\times 10^{-1}$ & $0.50$ &
$15.25$ & \\ $^{8}{\rm B}$ & $2.69\times 10^{-4}$ & $1.79$ & $3.74$ &
$1.84$\\ $^{13}{\rm N}$ & $3.13\times 10^{-2}$ & $0.03$ & $1.11$ & \\
$^{15}{\rm O}$ & $2.56\times 10^{-2}$ & $0.11$ & $1.68$ & \\ \hline &
& $2.56^{+0.40}_{-0.33}$ & $65.05^{+4.03}_{-3.06}$ &
$1.84^{+0.36}_{-0.26}$\\ \hline & & $2.56\pm 0.23$ & $75.6\pm 4.8$ &
$1.75 \pm 0.14$\\ \hline
\end{tabular}\\

\end{document}